\documentclass[prodmode,acmtoms]{acmsmall}

\usepackage[font=footnotesize]{caption}
\usepackage{amssymb}
\usepackage{graphics}
\usepackage[ruled, vlined]{algorithm2e}
\usepackage{sepnum}
\usepackage[mathcal]{euscript}
\usepackage{amsmath} 
\usepackage{amsfonts}
\usepackage{subfig}
\usepackage{tabularx}
\usepackage{url}

\def \Oh{{\it O\/}}
\newcommand{\oomit}[1]{}

\begin{document}

\markboth{N. Zhang, A. Roux  and T.  Zastawniak}{Parallel Binomial American Option Pricing with (and without) Transaction Costs}

\title{Parallel Binomial American Option Pricing with (and without) Transaction Costs}
\author{NAN~ZHANG
\affil{Xi'an Jiaotong-Liverpool University}
ALET~ROUX
\and
TOMASZ~ZASTAWNIAK
\affil{The University of York}}

\begin{abstract}
We present a parallel algorithm that computes the ask and bid prices of an American option when proportional transaction costs apply to the trading of the underlying asset. The algorithm computes the prices on recombining binomial trees, and is designed for modern multi-core processors. Although parallel option pricing has been well studied, none of the existing approaches takes transaction costs into consideration. The algorithm that we propose partitions a binomial tree into blocks. In any round of computation a block is further partitioned into regions which are assigned to distinct processors. To minimise load imbalance the assignment of nodes to processors is dynamically adjusted before each new round starts. Synchronisation is required both within a round and between two successive rounds. The parallel speedup of the algorithm is proportional to the number of processors used. The parallel algorithm was implemented in C/C++ via POSIX Threads, and was tested on a machine with 8 processors. In the pricing of an American put option, the parallel speedup against an efficient sequential implementation was 5.26 using 8 processors and 1500 time steps, achieving a parallel efficiency of 65.75\%. 
\end{abstract}




\acmformat{Zhang N., Roux A. and Zastawniak T. 2011. Parallel Binomial American Option Pricing with (and without) Transaction Costs}

\begin{bottomstuff}
Author's addresses: N. Zhang, Department of Computer Science and Software Engineering, Xi'an Jiaotong-Liverpool University, Suzhou, China; email:nan.zhang@xjtlu.edu.cn; Y. Roux  {and} T.  Zastawniak, Department of Mathematics, The University of York, York, UK; email:\{alet.roux, tomasz.zastawniak\}@york.ac.uk.
\end{bottomstuff}

\maketitle

\section{Introduction}
\label{section:introduction}
An American call (put) option is a financial derivative contract which gives the option holder the right but not the obligation to buy (sell) one unit of a certain asset (stock) for the exercise price $K$ at any time until a future expiration date $T$.  Option pricing is the problem of computing the price of an option, and is crucial to many financial practices. Since the classic work on this topic by \citeN{pricing-fischer-1973} and \citeN{theory-robert-1973}, many new developments have been introduced. In this paper, we present a parallel algorithm and its multi-threaded implementation that computes the ask and bid prices of an American option when proportional transaction costs apply to the underlying asset trading. Previous work on parallel valuation of European and/or American options can be found in \cite{architecture-alexandros-2004,parallel-alexandros-2010,programming-anwar-2007,parallel-ying-2010,option-steven-2010,high-mohammad-2008}, but they all assumed zero transaction cost in the underlying asset trading, which is often not the case.

When the underlying transaction costs are considered, the no-arbitrage price of an American option is no longer unique, but is confined within an interval. The upper bound of this interval is the ask price of the option, and the lower bound the bid price. For an American option based on a single underlying asset, its ask price can be derived from Algorithm 3.1 in \cite{american-alet-2009}, and its bid price from Algorithm 3.5. Unlike the previous approaches \cite{derivative-stylianos-1997,option-stylianos-2000,american-stylianos-2004,option-phelim-1992,derivative-bernard-1992} in pricing American/European options under transaction costs, the applicability of Algorithms 3.1 and 3.5 is not confined by the values of certain market and model parameters, or by methods of settlement (cash or physical delivery of the underlying asset). Besides pricing vanilla options such as puts and calls, the algorithms can also be applied to the valuation of options with more complex payoffs, such as American bull spreads.

The parallel algorithm that we present in this paper computes the ask and bid prices on recombining binomial trees, and was implemented in C/C++ via POSIX Threads. The implementation was tested on a machine with 8 processors (2 sockets $\times$ quad-core Intel Xeon E5405 at 2.0GHz). Experimental results showed that, for example, when the number $N$ of time steps was 1500 the parallel speedup in pricing an American put option was 5.26. Compared to the results obtained in the previous work \cite{architecture-alexandros-2004,parallel-alexandros-2010,parallel-ying-2010} this multi-threaded approach reduces the overhead of parallelisation and gains speedups on problems of much smaller sizes.

The contributions of this work are twofold. First, a parallel algorithm is designed and implemented which computes the ask and bid prices of American options under proportional transaction costs, whereas previous work for the same problem did not take transaction costs into consideration. Second, a refined generic strategy for partitioning a recombining binomial tree is developed. Like the previous partition schemes \cite{architecture-alexandros-2004,parallel-alexandros-2010,parallel-ying-2010,high-mohammad-2008}, our algorithm divides the whole tree into blocks consisting of nodes from multiple levels (where each level in the binomial tree consists of nodes at a particular time step). Each of these blocks is further divided into regions which are assigned to distinct processors in each single round of the computation. The previous schemes fixed each processor's assignment from the start of the computation. However, as the computation proceeds towards the root of the binomial tree the parallelism that can be exploited decreases. So, with a fixed assignment the load imbalance between different processors becomes more severe as the computation progresses. However our partition scheme re-calculates each processor's workload before the start of each new round so as to minimise the imbalance.  The partition scheme is generic in the sense that its applicability is not confined by the choice of the parameter values. 

The parallel binomial algorithm we developed is not specific to this particular problem of pricing American options under transaction costs. In the appendix we show the application of this parallel algorithm in pricing American options without transaction costs. The source codes for both these two applications of the parallel binomial algorithm are freely available through email\footnote{\url{nan.zhang@xjtlu.edu.cn}}.

\paragraph{Organisation of the rest of the paper.} Related work is reviewed in Section~\ref{section:related}. The sequential pricing algorithms are briefly explained in Section \ref{section:sequential}. The parallel algorithm and its analysis are presented in Section \ref{section:parallel}. Experimental results are reported in Section \ref{section:testing}. Conclusions are drawn in Section \ref{section:conclusion}, which also contains a discussion of future work. The appendix contains a discussion about applying the parallel algorithm to the pricing of American options with no transaction costs, and presents the results from the performance tests on the same machine.

\section{Related work}
\label{section:related}
Previous approaches in parallel option pricing are discussed in this section. None of this work took transaction costs into consideration.

To exploit data-parallelism on recombining binomial/trinomial trees, a parallel option pricing algorithm must partition a whole tree into blocks and assign them to distinct processors for parallel processing. Some approaches \cite{architecture-alexandros-2004,parallel-alexandros-2010,parallel-ying-2010,high-mohammad-2008} divided a binomial/trinomial tree into blocks consisting of multiple levels of nodes, and processed the blocks using multiple processors. But some \cite{options-Craig-2005,option-steven-2010} processed nodes of a single level in parallel and afterwards moved to the next-highest level in sequential order. Compared with the latter method, the former requires more sophisticated synchronisation strategies and thus is more complicated to implement. But its advantage is that it causes less parallelisation overhead. The partition scheme we designed in our algorithm belongs to the first class.

\citeN{architecture-alexandros-2004} presented an architecture-independent parallel pricing algorithm for American and European-style options on recombining binomial trees. The algorithm partitioned a binomial tree into $b \times b$ blocks and assigned these blocks to distinct processors in a wrapped-mapping manner such that the maximum input data imbalance between any two processors is limited by~$b$. This assignment (Fig. 5 in \cite{architecture-alexandros-2004}) was determined from the start of the computation according to the number of leaf nodes at level $N$ and the number $p$ of processors involved. The computation on the whole binomial tree was divided into rounds, where in each round $b$ levels of the tree were processed. No load re-balancing was applied after each round of the computation. The parallelisation was achieved via the Oxford BSP (Bulk Synchronous Parallel) \cite{parallel-rob-2004} Toolset, BSPlib, and another non-BSP message passing interface (MPI) LAM-MPI \cite{greg-lam-1994}. The implementation was tested on a cluster of 16 PC workstations, each running a dual-Pentium 350 MHz. Their tests showed that when $N = 8192$ and $b = 128$, using the BSPlib, the parallel speedup was 2.71 when $p = 8$ and 3.19 (Table 1 in \cite{architecture-alexandros-2004}) when $p = 16$. When implemented via the LAM-MPI, the speedup was 2.23 and 2.28 (Table 5 in \cite{architecture-alexandros-2004}), respectively.

\citeN{parallel-ying-2010} presented a parallel option pricing algorithm based on a Backward Stochastic Differential Equation (BSDE). The computation was performed on binomial trees that model the Brownian dynamic change of the underlying asset price. The algorithm assumed the number $N$ of time steps and the number $p$ of processors to be a power of two. To avoid frequent communications they introduced a parameter $L$ such that in each iteration of the computation $L$ levels of nodes were processed in parallel. Their algorithm assumed that $L$ was a power of two plus one and $N$ was divisible by $L - 1$. Each processor's assignment (Fig. 2 in \cite{parallel-ying-2010}) was fixed at the start of the computation. No load re-balancing was attempted afterwards. The algorithm was implemented in C via MPI. Tests were made on a cluster of 16 PC nodes where each node ran 2 Intel Xeon DP 2.87 GHz. The parallel speedup was 3.15 using 8 processors and 3.33 (Table 1 in \cite{parallel-ying-2010}) using 16, when $N = 8192$ and $L = 9$. 

A GPU-based (graphics processing unit) solution \cite{parallel-bin-2010} to the BSDE approach for option pricing was presented by the same group of researchers, where they adopted the theta method \cite{new-weidong-2006} to solve BSDEs. (The theta method discretises a continuous BSDE on a time-space grid. At each node of the grid Monte Carlo simulations are used to approximate the mathematical expectations. The whole process requires a large amount of calculations but suits the computing architecture of a GPU.) The implementations were tested on a 2.67 GHz Intel Core i7 920 and an NVIDIA Tesla C1060. When $N = 128$ the runtime of the sequential code was about 23000 seconds, and that of the GPU code was about 99 seconds (Table 1 in \cite{parallel-bin-2010})!

\citeN{high-mohammad-2008} proposed a cache-friendly parallel option pricing algorithm for shared memory symmetric multi-processors (SMP). The algorithm gave much consideration to the memory hierarchy available in modern RISC processors. In order to be cache-efficient the algorithm employed techniques such as cache and register blocking, and partitioned a binomial tree into triangular and quadrangular blocks (Fig.~8 in \cite{high-mohammad-2008}). As the computation proceeded towards the root of the tree the number of blocks decreased and so did the number of processors that could be utilised. The algorithm was implemented in Fortran 95 with parallelisation achieved via OpenMP directives \cite{rajat-techniques-2001}. A test of the parallel algorithm on 8 Sun UltraSPARC III 1050MHz processors showed that when the block size was 128, $N = 8192$ the parallel speedup was 4.96 (Table 4 in \cite{high-mohammad-2008}) using all the 8 processors. A similar serial cache-friendly option pricing algorithm was discussed by \citeN{cache-john-2010}. It was based on the binomial and trinomial models without parallelisation of any type.

As a supplement to the latency-tolerant BSP-oriented algorithms for option pricing on binomial trees in \cite{architecture-alexandros-2004} and trinomial trees in \cite{alexandros-trinomial-2003}, \citeN{parallel-alexandros-2010} presented a more up-to-date parallel algorithm using the explicit finite difference method \cite{valuation-eduardo-1977}, which is equivalent to computing discounted expectations on a trinomial tree. The algorithm partitioned the nodes of a trinomial tree into rectangular blocks (Fig. 4 in \cite{parallel-alexandros-2010}) of $b$ levels. As in \cite{architecture-alexandros-2004}, the nodes of a block were further divided into three regions, one for nodes for which the discounted expectations have already been computed, one for nodes for which the computation does not depend on the results from nodes in a neighbouring block, and one for nodes for which such dependency exists. The algorithm was implemented via the Oxford BSP Toolset, the non BSP-specific libraries LAM\_MPI and Open MPI \cite{open-richard-2005}, and SWARM \cite{david-swarm-2007}, a parallel computing framework for multi-core processors. Their tests were done on the same PC cluster as in \cite{architecture-alexandros-2004} and on two multi-core processors. On the 2.4GHz Intel quad-core Q6600 used in their tests, the parallel speedup was 3.63 using BSP and MPI, and 3.13 (Table 11 in \cite{parallel-alexandros-2010}) using SWARM when $N = 8192$, $b = 129$ and $p = 4$.

\citeN{programming-anwar-2007} published a white paper where parallel binomial option pricing was implemented in Ct\footnote{After Intel's merge with RapidMind technologies, Ct became a part of what is now known as the Intel Array Building Blocks (ArBB).}, a data parallel API implemented within the C++-based syntactic framework. The parallel code was tested on two 2.33GHz Intel Xeon quad-core E5345, and gained much speedup over a sequential C++ implementation thanks to Ct's built-in SSE-based implementation for the common math functions.

\citeN{option-steven-2010} presented a GPU-based parallel solution for pricing American lookback options on recombining binomial trees. The algorithm did backward computation on a binomial tree with nodes at each level being processed in parallel. Initially, the computation was carried out by the GPU, but after a certain threshold level was passed the computation was taken over by the CPU, because as the calculation proceeded to the root of the tree the parallelism that could be exploited decreased. Their tests were performed on a 3.0GHz Intel Core 2 Duo and a 216-core NVIDIA GTX 260. The speedup of the CPU+GPU hybrid implementation against an un-optimised sequential code was about 20 (Fig. 7 in \cite{option-steven-2010}) when the number of time steps was 5000 and the threshold was set as 256. The same partition scheme was used by the GPU-based parallel binomial option pricing discussed by \citeN{options-Craig-2005}, where the nodes in each single level of a tree were processed in parallel.

\citeN{parallel-kai-2005} presented a parallel algorithm for pricing basket American-style Asian options on recombining binomial trees. The number of levels in a tree and the number of processors were assumed to be a power of two. To partition a tree, initially, all leaf nodes were evenly distributed among the processors. The computation proceeded to the root of the tree in such a way that in a given processor for every pair of adjacent nodes at a certain level $i$ the processor computed the option price for the pair's parent node at level $i - 1$ (Fig.~3 in \cite{parallel-kai-2005}). Eventually processor 0 computed the option price at level 0. No load re-balancing among the processors was attempted during the course of the computation. The implementation of the algorithm was in C via MPI.

Compared with these parallel approaches in binomial option pricing, the generic partition scheme in our algorithm makes ample allowance for minimising the load imbalance between processors to enhance the efficiency of the parallelisation. The multi-threaded implementation of the algorithm is light-weight: the parallel speedup on 8 processors in a tested American put option is 5.26 when $N = 1500$.

Algorithms for parallel option pricing based on models other than the binomial/trinomial tree can be found as well. These are loosely connected to what we present in this paper. \citeN{option-gianluca-2010} published a numerical procedure for pricing exotic path-dependent options when the underlying asset price evolves according to a generic L\'evy process \cite{levy-wim-2003}. By geometric randomisation of the option expiration, the $n$-step backward recursion in option pricing was transformed into an integral equation. The option price was then obtained by solving $n$ independent integral equations. Because the equations were mutually independent they were solved in parallel on a grid computing architecture.

\citeN{parallel-vladimir-2010} presented algorithms based on the Fourier space time-stepping method to price single- and multi-asset European and American options with stock prices following exponential  L\'evy processes. The algorithms were implemented on an NVIDIA GeGorce 9800 GX2 video card with only one of the two GPUs being used.

\citeN{parallel-hari-2010} proposed a parallel synchronous option pricing algorithm to price simple European options using particle swarm optimisation: a nature-inspired global search algorithm based on swarm intelligence.

\citeN{halis-parallel-2007} discussed the application of parallel computing in pricing backward-starting fixed strike Asian options that are continuously averaged. Through a change of numeraire they transformed the pricing problem into solving a one-state-variable partial differential equation (PDE) by both explicit and Crank-Nicolson's implicit finite-difference methods. The algorithms they designed were implemented via MPI and were tested on a Linux PC cluster.

\section{The sequential pricing algorithms}
\label{section:sequential}
\begin{figure}[t]

\centering

\subfloat[Without transaction costs.]{\includegraphics{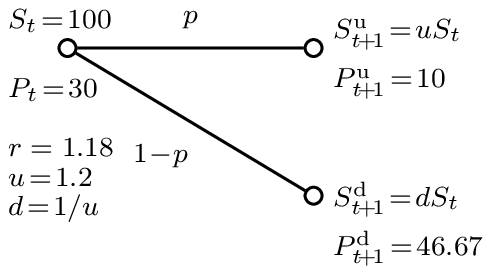}\label{subfigure:noCosts}}
\hspace{0.3cm}
\subfloat[With transaction costs.]{\includegraphics{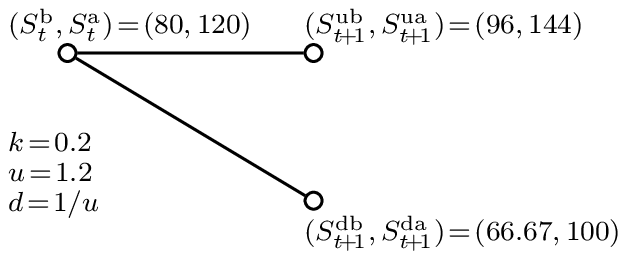}\label{subfigure:withCosts}} 
\caption{One-step binomial processes with and without transaction costs.}
\label{figure:one-step}
\end{figure}

We first briefly go through the idea of pricing American options when transaction costs are not included. Consider an American put option with strike~$K$ and expiry $T$, which can be exercised once at any time $0, 1, 2, \ldots, N$. We use the one-step binomial process example in Fig. \ref{figure:one-step}\subref{subfigure:noCosts}, where at time $t$ the price of the underlying stock of an American put option is $S_t$. After the one time step, the price of the stock can either be $uS_t$ or $dS_t = u^{-1}S_t$.  We assume the interest rate over the one step time period is $\rho$, that is, 1 unit of cash bond at time $t$ will grow to $r=1+\rho$ units at time $t + 1$. The risk-neutral probability for the up-move is $p = (r - d) / (u - d)$, and for the down-move is $1 - p$. The payoff of the American option at time $t$ is $P_t = \max(K - S_t, 0)$; that is, if $S_t<K$ then the owner of the option will exercise his/her right to sell one unit of the stock (worth $S_t$) at the price $K$, thus making a profit of $K-S_t$, and if $S_t\ge K$ then exercising the option is not advantageous. The option is priced by backward induction, which gives a unique arbitrage-free price $\pi_t$ for the option at time~$t$. At the maturity date $N$ the value of the option is the same as its payoff, so $\pi_N=P_N$. For $t<N$, the value $\pi_t$ of the option at the node $S_t$ is the maximum of its discounted expected payoff $r^{-1} \mathbb{E}(\pi_{t\!+\!1}|S_t) = r^{-1} (p\pi_{t\!+\!1}^{\mathrm{u}} + (1-p)\pi_{t\!+\!1}^{\mathrm{d}})$ at time $t$ and its immediate payoff $P_t$ if the option is exercised at $t$, that is $\pi_t = \max(P_t, \mathbb{E}(\pi_{t\!+\!1}|S_t)/r)$. For example, $p=0.9454$ for the parameter values in Fig.~\ref{figure:one-step}\subref{subfigure:noCosts}, which also shows the option payoffs for $K=130$. Now suppose that the option prices at the nodes at time $t+1$ have already been computed and happen to coincide with the corresponding payoffs, $\pi^{\mathrm{u}}_{t\!+\!1}=10$ and $\pi^{\mathrm{d}}_{t\!+\!1}=46.67$ (that is, in this example both these nodes are in the exercise region for the American option). Then we can compute $\pi_t=\max(30,10.17)=30$. To compute $\pi_0$ on a binomial tree of multiple levels, we start from the leaf nodes and go all the way back to the root to obtain the price of the option at time 0.

Proportional transaction costs in asset (stock) tradings are modelled by bid-ask spreads. That is, at time $t$ a unit of stock can be bought for the ask price $S^\mathrm{a}_t$ or sold at the bid price $S^\mathrm{b}_t$. To link this with the friction-free model, we shall assume that $S^\mathrm{a}_t=(1+k)S_t$ and $S^\mathrm{b}_t=(1-k)S_t$, where $k\in[0,1)$ is the transaction cost rate. Under these conditions the arbitrage-free price of an American option at any time $t$ is no longer unique, but is confined within an interval. The upper limit of this interval is the ask price $\pi^{\mathrm{a}}_t$ of the option, and the lower limit the bid price $\pi^{\mathrm{b}}_t$. The ask price is the price at which the option can be bought on demand. It is also the minimum amount of wealth that the seller of the option needs to hedge his/her positions in all circumstances, that is, to deliver to the buyer the obligated payoff portfolio without having to inject extra wealth. The bid price is the price at which the option can be sold on demand. It is also the maximum amount of wealth that the buyer can borrow against the right to exercise the option.

Let $(\xi, \zeta)$ be the payoff process of an American option, that is, if the holder exercises the option at any time $t = 0, 1, 2, \ldots, N$, then the seller must deliver to the holder a portfolio consisting of $\xi_t$ cash and $\zeta_t$ units of the asset (stock). For the above American put option this is $(K, -1)$ at all times. To hedge his/her position the seller should hold a portfolio consisting of cash and the underlying stock, and we use $(x_t, y_t)$ to denote his/her holdings of cash and stock at time $t$. We define the seller's expense function $u_t$ at time $t$ to be 
\begin{equation}
u_t(y) = \xi_t + (y - \zeta_t)^- S_t^{\mathrm{a}} - (y - \zeta_t)^+ S_t^{\mathrm{b}}, 
\end{equation}
where $(y - \zeta_t)^- = -\min(y - \zeta_t, 0)$ and $(y - \zeta_t)^+ = \max(y - \zeta_t, 0)$. This is a function of the seller's stock holding at time $t$. It defines the minimum amount of cash that the seller needs at $t$ to fulfil his/her obligation if the option is exercised at $t$. So if the seller wishes to form a self-financing strategy to cover his/her position at $t$, his/her holdings $(x_t, y_t)$ must belong to the epigraph of $u_t$, that is $(x_t, y_t) \in \mathrm{epi}\, u_t$. (The epigraph of any function $f$ is the set of points which lie above $f$ in the $yx$-plane, namely $\mathrm{epi}\, f = \{(y, x) \in \mathbb{R}^2\ |\ x \geq f(y)\}$.)

\begin{figure}

\centering
\includegraphics{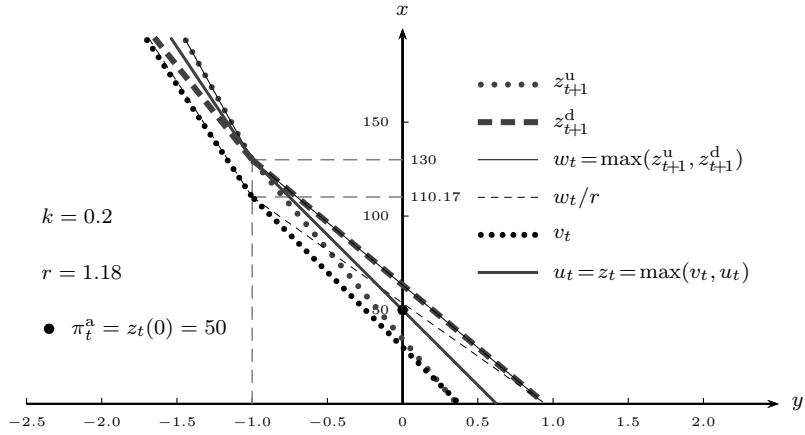}
\caption{The piecewise linear functions in computing the ask price.}
\label{figure:seller}
\end{figure}

Now using the same American put example (with $K = 130$) and the one-time step binomial process (Fig. \ref{figure:one-step}\subref{subfigure:withCosts}), we explain how the option ask price $\pi^{\mathrm{a}}_t$ is computed at any time $t$. This is done by constructing a sequence of piecewise linear functions $z_t$ by backward induction from time step~$N$, when $z_N=u_N$. The interpretation of $z_t$ is that a portfolio $(x,y)$ at time $t$ allows the seller to deliver the option without risk if and only if $(x,y)\in\mathrm{epi}\, z_t$. For $t<T$, we start from the two nodes at time $t+1$. Suppose that 

\begin{equation}
z_{t\!+\!1}^{\mathrm{u}}(y) = u_{t\!+\!1}^{\mathrm{u}}(y) = 130 + 144(y+1)^- - 96(y+1)^+
\end{equation}

\noindent at the up-move node. This is a piecewise linear function because

\begin{equation}
z_{t\!+\!1}^{\mathrm{u}}(y)  = \left\{ \begin{array}{lp{1cm}l} -144y - 14 & & y < -1  \\  -96y + 34 & & y \geq -1 \end{array} \right.;
\end{equation}

\noindent see Fig. \ref{figure:seller}. For the down-move node suppose that 

\begin{equation}
z_{t\!+\!1}^{\mathrm{d}}(y) = u_{t\!+\!1}^{\mathrm{d}}(y) = 130 + 100(y+1)^- - 66.67(y+1)^+.
\end{equation}

At time $t$, because the seller must be prepared for the worst case scenario, we calculate the maximum of $z_{t\!+\!1}^{\mathrm{u}}$ and  $z_{t\!+\!1}^{\mathrm{d}}$, to obtain $w_t = \max(z_{t\!+\!1}^{\mathrm{u}}, z_{t\!+\!1}^{\mathrm{d}})$. Now since $x$ units of cash at time $t$ will grow to $xr$ at time $t+1$, the function $w_t$ must be discounted by $r$. Now the slopes of this discounted function $w_t / r$ must be restricted within the interval $[-S_t^{\mathrm{a}}, -S_t^{\mathrm{b}}]$, which is $[-120, -80]$ in this example, to account for the possibility of rebalancing the portfolio at time~$t$. This restricted function is denoted by $v_t$. It is the discounted expected expense function at time $t$. The epigraph of this function consists of portfolios covering the option seller if the option is exercised at time $t+1$ or later. Now what if the option is exercised at time $t$? The expense function $u_t$ at $t$ is $u_t = 130 + 120(y + 1)^- - 80(y + 1)^+$. Again, the seller must be prepared for the worst, which corresponds to the expense function being the maximum\begin{equation}
z_t(y) = \max(u_t(y), v_t(y)) = u_t(y) \left\{ \begin{array}{lp{1cm}l} -120y + 10 & & y < -1  \\  -80y + 50 & & y \geq -1 \end{array} \right..
\end{equation}
These piecewise linear functions are shown in Fig. \ref{figure:seller}. The option ask price at time $t$ for this example is then $\pi^{\mathrm{a}}_t = z_t(0) = 50$, because it is the minimum amount of cash that enables a seller without a stock holding to hedge his/her position without risk at time $t$. When the above computation is carried out on a binomial tree representing $N$ time steps, we start from the leaf nodes and work backwards to the root node at time 0. The option ask price is then $\pi^{\mathrm{a}}_0 = z_0(0)$.

\begin{figure}

\centering
\includegraphics{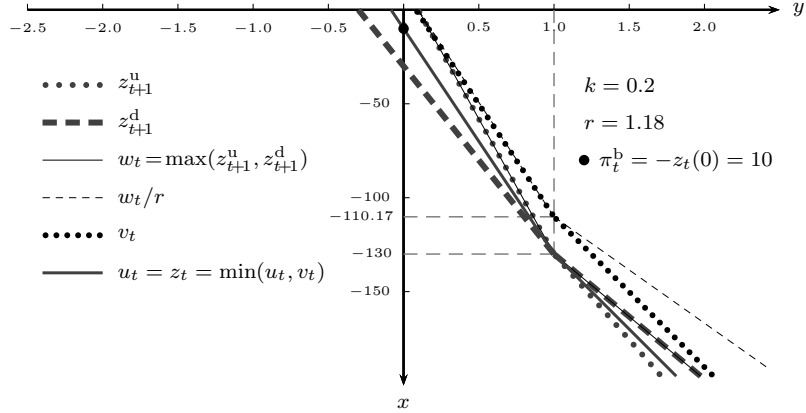}
\caption{The piecewise linear functions in computing the bid price.}
\label{figure:buyer}
\end{figure}

For the buyer's case, the buyer's expense function at time $t$ is 
\begin{equation}
u_t(y) = -\xi_t + (y + \zeta_t)^- S_t^{\mathrm{a}} - (y + \zeta_t)^+ S_t^{\mathrm{b}}, 
\end{equation}
because it is he/she who will receive the portfolio $(\xi_t, \zeta_t)$. The pricing procedure for $z_N$, $w_t$ and $v_t$ is similar to that for the seller. But when $z_t$ is computed the minimum operation is used instead of the maximum. The reason for this difference is that at any time $t<N$ the buyer needs to choose between exercising or waiting (and choose a portfolio in $\mathrm{epi}\, u_t$ or $\mathrm{epi}\, v_t$), whereas the seller needs to be prepared for any eventuality (i.e. they need a portfolio in $\mathrm{epi}\, u_t$ and $\mathrm{epi}\, v_t$). In this example, if it is assumed that
\begin{align*}
 z_{t\!+\!1}^\mathrm{u}(y) &= -130 + 144(y-1)^- - 96(y-1)^+, \\
 z_{t\!+\!1}^\mathrm{d}(y) &= -130 + 100(y-1)^- - 66.67(y-1)^+,
\end{align*}
then
\begin{equation}
z_t(y) = \min(u_t(y), v_t(y)) = u_t(y) = \left\{ \begin{array}{lp{1cm}l} -120y - 10 & & y < 1  \\  -80y - 50 & & y \geq 1 \end{array} \right..
\end{equation}
The option bid price $\pi^{\mathrm{b}}_t$ at time $t$ is $\pi^{\mathrm{b}}_t = -z_t(0) = 10$, because the bid price is the maximum amount of wealth that the buyer can borrow against the right of exercise. See Fig. \ref{figure:buyer} for a plot of the piecewise linear functions in the buyer's case.

Full details of the procedures for finding bid and ask prices under proportional transaction costs can be found in Algorithms 3.1 and 3.5 in \cite{american-alet-2009}. Note that in general $\pi_t \in [\pi^{\mathrm{b}}_t, \pi^{\mathrm{a}}_t]$. 

\section{The parallel pricing algorithm}
\label{section:parallel}
\subsection{Binomial tree model}
For an American option whose payoff process and physical expiration time are $(\xi, \zeta)$ and~$T$, respectively, let $N$ be the number of time intervals that discretise the time period from 0 to $T$. Also let $\sigma$ be the volatility of the underlying stock, $R$ the continuously compounded annual interest rate and $k \in [0, 1)$ the transaction cost rate. Under such conditions the binomial tree that models the dynamics of the stock price will have $N + 1$ levels, corresponding to the time steps $t = 0, 1, 2, \ldots, N$. The up-move factor $u$, down-move factor $d$ and cash accumulation factor $r$ over one time step are $u = \exp(\sigma \sqrt{T/N})$, $d =u^{-1} = \exp(-\sigma \sqrt{T/N})$, and $r = \exp(RT/N)$, respectively. The pricing algorithms in \cite{american-alet-2009} actually add an extra time instant $t = N + 1$ to the model and set the option payoff as $(0, 0)$ at all the $N + 2$ nodes in that level. The purpose of adding this extra time step is to model the possibility that under certain circumstances it may be in the best interests of the option holder to leave it unexercised. In line with \cite{american-stylianos-2004} and \cite{american-alet-2009}, we assume that no transaction costs apply at time 0, that is, $S_0^{\mathrm{b}} = S_0 = S_0^{\mathrm{a}}$.

\subsection{The partition scheme and the synchronisation mechanism}
Assume we have $p$ distinct processors in a parallel computer. Because the computation of the $u, w, v, z$ functions at different nodes can be performed independently in parallel, we can partition a whole binomial tree into blocks of nodes and assign these blocks to distinct processors. The parallel algorithm, like its sequential counterpart, starts off at the leaf nodes where $t = N + 1$ and works backwards towards the root of the tree. The whole process is accomplished by $p$ threads, denoted by $p_0, p_1, \ldots, p_{p\! -\! 1}$, with each thread being bound explicitly to a distinct processor. The whole computation is divided into rounds, where in each round the nodes of a block are processed by the $p$ threads in parallel. 

In general, if the base level $B$ (whose nodes have been processed in the $(i-1)$th round) of an $i$th round is at time $t = n, n \in [1, N+1]$, then the total number of nodes in at that level will be $n + 1$. These $n+1$ nodes will be divided equally among the $p$ threads. So all the threads $p_i, i = 0, 1, \ldots, p\! -\! 2$, get $\lfloor (n+1) / p\rfloor$ nodes, but the last thread $p_{p\! -\! 1}$ gets $(n+1) - \lfloor (n+1)(p-1) / p\rfloor$ nodes. We use $L$ to denote the maximum number of levels that are processed towards the root in a round, that is, the maximum number of levels in a block. However, the number $D$ of levels that are actually processed in a round is jointly determined by $L$ and the number of nodes that each thread gets, because this number $D$ cannot exceed $\lfloor (n+1) / p\rfloor - 1$. So we have $D = \min(L, \lfloor (n+1) / p\rfloor - 1)$. So in a round whose base level $B$ contains $n + 1$ nodes all the threads will be assigned a block of $\lfloor (n+1) / p\rfloor \times D$ nodes, except the last thread $p_{p\! -\! 1}$ which only gets a smaller number of nodes. For a thread $p_i, i \in [0, p-2]$, we further divide its $\lfloor (n+1) / p\rfloor \times D$ nodes into region A and region B such that the computations performed at the nodes in region A do not depend on the results from another thread in the same round, but the computations at the nodes in region B do need results from thread $p_{i\!+\!1}$. Note that the last thread $p_{p\! -\! 1}$ does not have any B nodes in any round of the computation. 

\begin{figure}[t]

\parbox[b]{.74\linewidth}{\includegraphics{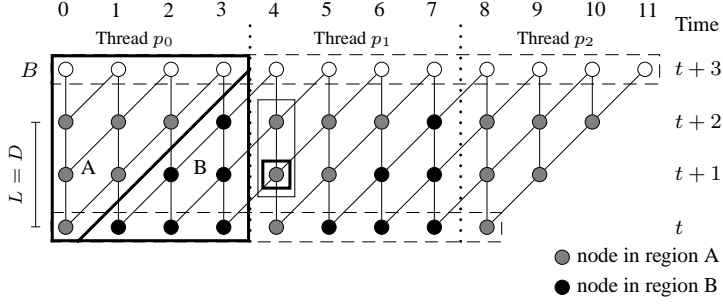}}\hfill
\parbox[b]{.26\linewidth}{

\caption{Partition into regions A and B.}
\label{figure:roundParition}
}
\end{figure}

Fig. \ref{figure:roundParition} shows such a division among 3 threads in a round consisting of 3 ($L = D = 3$) levels of nodes. The nodes enclosed by the dashed frame box at time level $t + 3$ are the base nodes. For thread $p_0$, to compute the $u, w, v, z$ functions at the nodes in its region B, it needs results computed by thread $p_1$ at the two nodes in column 4 enclosed by the thin frame box. Thread $p_0$ cannot start computing at the nodes in its region B until thread $p_1$ finishes at the node (level $t + 1$, column 4) enclosed by the bold frame box. In general, thread $p_i$, $i \in [0, p-2]$, cannot start at nodes in its region B until thread $p_{i\!+\!1}$ finishes at the leftmost node at level $B-D+1$ in its region A. This scheme of partitioning into A and B regions was also adopted in \cite{architecture-alexandros-2004} and \cite{parallel-ying-2010}.

\begin{figure}

\parbox[b]{.74\linewidth}{\includegraphics{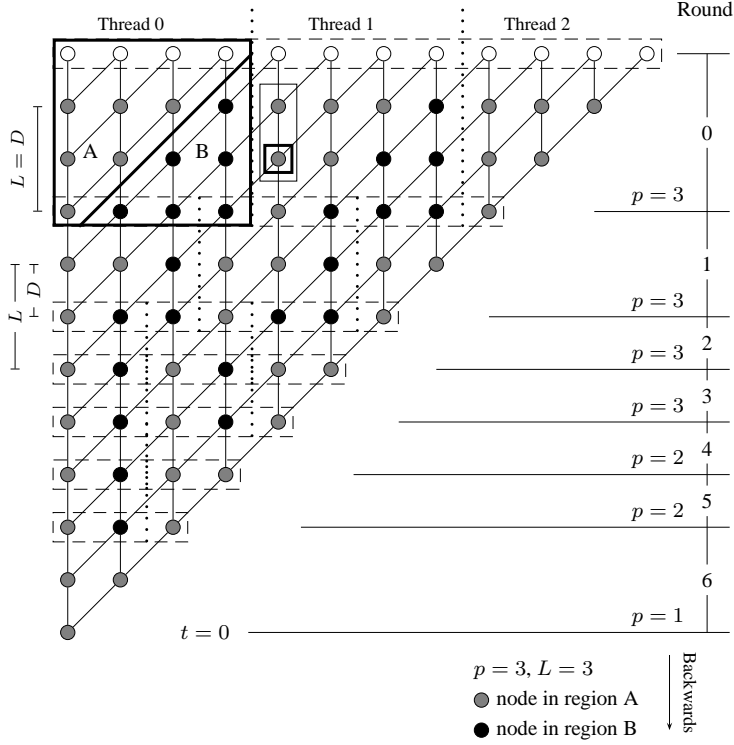}}\hfill
\parbox[b]{.26\linewidth}{

\caption{Parallel processing on the binomial tree by three working threads. Note that although $N=10$ the algorithm adds an extra time instant to the model.}
\label{figure:tree}
}
\end{figure}

The parallel algorithm re-balances the workload of each thread after each round of the computation. If the current base level is $B$, the next base level will be $B-D$, containing $B-D+1$ nodes, and according to this number the workload of each thread in the next round will be calculated. The parallel algorithm ensures that each thread will get minimally two nodes to process in all the rounds, which means that the minimum possible value that $D$ can get is 1. If at some level of the tree the number of nodes is less than $2p$ the number of processors used will be decreased by 1 until this no-less-than-two-node condition is satisfied. A partition based on the above explanation is shown in Fig. \ref{figure:tree} for $N=10$, $p=3$ and $L=3$. The figure shows the adjustment of the workload after each round and the reduction in the number of processors needed as the computation proceeds towards the root of the tree.

To save the intermediate $z$ functions generated during the computation, instead of generating the whole tree, the parallel algorithm maintains two buffers, each with $(L+1)$ rows $\times$ $(N+2)$ columns. One of these two buffers is for computing the ask price, and the other the bid price. The mapping between a whole binomial tree and the buffers is done in a modular wrapping around manner to avoid the cost of extra synchronisation and copy back. We use variable $U$ to denote the base level in the two buffers in a round of the computation, corresponding to the base level of the tree. Initially, this $U$ is set to 0, and after a round whose base level of the tree is $B$ and works $D$ levels towards the root, $U$ is updated by $U \leftarrow (U + (B - D))\! \mod\! (L + 1)$. Now suppose the computation is working on the $i$th, $i \in [0, D]$, level down from level $B$. The piecewise linear functions will be computed and stored in the two buffers at level $(U + i)\! \mod\! (L + 1)$ according to the piecewise linear functions stored at level $(U + i - 1)\! \mod\! (L + 1)$ in the buffers.

\begin{figure}[t]


\parbox[b]{.7\linewidth}{\includegraphics{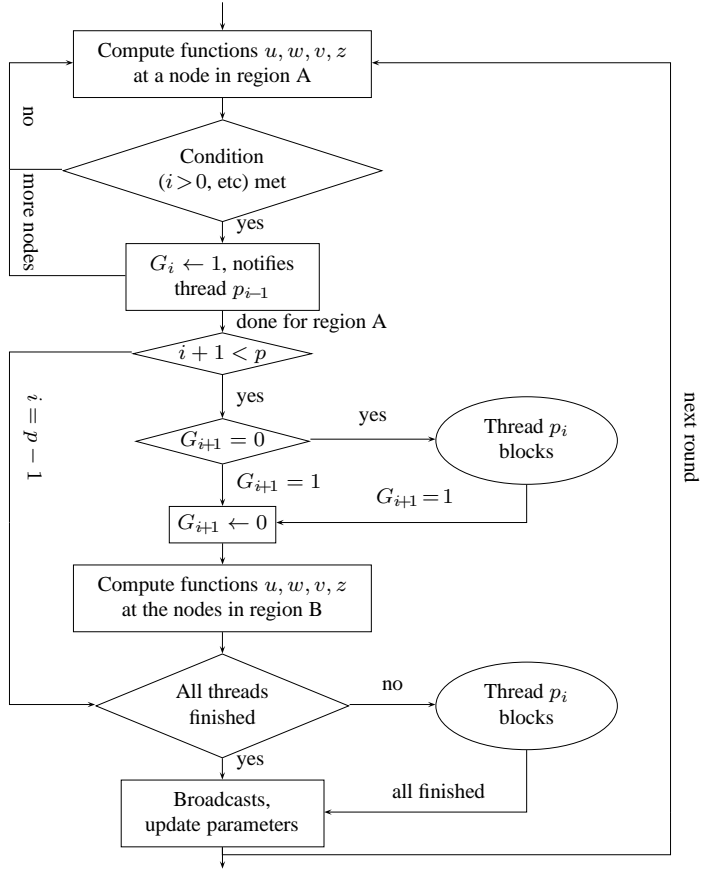}}\hfill
\parbox[b]{.3\linewidth}{
\caption{The synchronisations on thread $p_i, i \in [0, p-1]$. The condition in the first rhombus box is shown at line 15 in Algorithm \ref{algorithm:threadi}.}
\label{figure:controlFlow}
}
\end{figure}

As the whole computation is divided into rounds, the threads have to be synchronised both within a round and between two successive rounds. Within a round, all the threads work $D$ levels down the tree in such a way that any two adjacent threads have to be synchronised. As soon as thread $p_i, i \in [1, p-1]$ has finished the leftmost node (such as the single node enclosed by the bold frame in Fig. \ref{figure:tree}) at level $B - D + 1$ ($B$ being the base level of the round) in its region A, it will send a signal $G_i$ to thread $p_{i\! -\! 1}$, so that after thread $p_{i\! -\! 1}$ has finished the nodes in its region A, upon receiving the signal it can proceed to the nodes in its region B. Once thread $p_i, i \in [1, p-1]$ has finished processing all its nodes in regions A and B, it has to wait for the other peer threads to finish their work. Only after all the threads have finished, can the parameters be updated for the next round. The flow chart in Fig. \ref{figure:controlFlow} using thread $p_i, i \in [0, p-1]$ as an example shows the synchronisation scheme.

The pseudo code in Algorithm \ref{algorithm:threadi} shows the computational steps performed by thread $p_i, i \in [0, p-1]$, including the synchronisation scheme. Note that node $(l, c)$ there denotes the node at level $l$ of the tree whose column index is $c$. The nested for-loop that computes the functions at nodes in region B is similar to the one in region A, so the details are omitted. The pseudo code is executed by all the threads $p_i, i = 0,1,2, \ldots, p-1$. Because thread $p_0$ is the one that computes $z_0$ at the root node at $t = 0$, the option ask and bid prices are returned by thread $p_0$. We finally have $\pi_0^{\mathrm{a}} = z_0(0)$ and $\pi_0^{\mathrm{b}} = -z_0'(0)$, where $z_0$ is the seller's expense function at $t=0$ and $z_0'$ the buyer's.

\begin{algorithm}[!t]
\begin{footnotesize}
\linesnumbered
\SetAlCapFnt{\footnotesize} 
\SetKwInput{Input}{Input}
\SetKwInput{Output}{Output}
\SetKwFor{ForAllParallel}{forall}{in parallel do}{endfor}

\Input{Up-move factor $\mathrm{u}$, interest rate $r$, number $p$ of processors, number $N$ of time steps, stock price $S_0$ at root node, transaction cost rate $k$.}
\Output{The functions $u$, $w$, $v$ and $z$ for the seller and the buyer at each node.}
\BlankLine
\Begin{

\tcp{Initialisation at nodes in level $t = N+1$.}

$n \leftarrow N + 1$, $s \leftarrow i \times \lfloor (n+1) / p\rfloor$;

$e \leftarrow (i+1) \times \lfloor (n+1) / p\rfloor$ $\forall\, i \neq p-1$, or $e \leftarrow n+1$ when $i = p-1$;

\For{$l \leftarrow s \mathrm{;}$ $l < e \mathrm{;}$ $l \leftarrow l + 1$}{
  $z_{N\! +\! 1}^l \leftarrow u_{N\! +\! 1}^l$ with payoff (0,0) for both the parties at node $(N+1, l)$;
}
\vspace{0.2cm}

\tcp{Start to work backwards down to the root.}

$U \leftarrow 0$; \tcp{Used for the mapping from the tree to the buffers.}

\For{$B \leftarrow N+1 \mathrm{;}$ $B > 0\ \mathrm{and}\ i < p \mathrm{;}$}{
  $D \leftarrow \min(L, \lfloor(n+1)/p\rfloor - 1)$;

  $o \leftarrow 1$; \tcp{$o$ is the column offset for region A.}

  \lIf{$D > 1$}
      { 
        $T \leftarrow B - D + 1$;
      }  \lElse{
        $T \leftarrow B - D$;} \tcp{$T$ is the level on which the signal $G_i$ is triggered.}

\vspace{0.2cm}

 \tcp{Compute functions $u$, $w$, $v$ and $z$ at the nodes in region A.}

  \For{$C \leftarrow B - 1 \mathrm{;}\ C \ge B - D \mathrm{;}\ C \leftarrow C - 1$}{
    $m \leftarrow \min(e - o, C + 1)$;

    \For{$l \leftarrow s \mathrm{;}\ l < m \mathrm{;}\ l \leftarrow l + 1$}{
      Compute $z_C^l$ for both the parties at node $(C, l)$;

      \If{$i > 0\ \mathrm{and}\ C = T\ \mathrm{and}\ l = s$}{
        $G_i \leftarrow 1$, and signal the change to thread $p_{i\!-\!1}$;
      }
      }
    $o \leftarrow o + 1$;
  }

\vspace{0.2cm}

\tcp{Compute functions $u$, $w$, $v$ and $z$ at the nodes in region B.}

  \If{$i + 1< p$}{
    Block until signal $G_{i\! +\! 1}$ becomes 1;

    $G_{i\! +\! 1} \leftarrow 0$; \tcp{Prepare for the next round.}

    Compute the $z$ functions at nodes in region B from level $B-1$ to $B-D$ whose column indexes are within $[s, e-1]$;

  }

  Wait until all threads reach this point;

\vspace{0.2cm}

\tcp{Start to update the parameters for the next round.}

  $B \leftarrow B-D$, $U \leftarrow (U + D) \mod (L + 1)$;

  \If{$B > 0$}{
    $n \leftarrow B + 1$; \tcp{$n$ is the number of nodes at the next base level.}

    \While{$n < (2 \times p)$}{
      $p \leftarrow \max(p - 1, 1)$; 
    }

    $s \leftarrow i \times \lfloor (n+1) / p\rfloor$;
    
    $e \leftarrow (i+1) \times \lfloor (n+1) / p\rfloor$ $\forall\, i \neq p-1$, or $e \leftarrow n+1$ when $i = p-1$;
  }

}

}
\caption{Computational steps executed by thread $p_i, i \in [0, p-1]$.}
\label{algorithm:threadi}
\end{footnotesize}
\end{algorithm}

\subsection{Computational time analysis}
Algorithms 3.1 and 3.5 in \cite{american-alet-2009} have polynomial runtime $T_{\mathrm{S}} = \Oh(N^k)$ for some $k \ge 2$. Although the number of nodes in a recombining binomial tree is quadratic in $N$ (so a traditional binomial pricing algorithm without transaction costs has runtime  $T_{\mathrm{S}}' = \Oh(N^2)$), the maximum, minimum and slope restriction operations may require slightly more time to finish as the computation proceeds towards the root because the piecewise linear functions $u$, $w$, $v$ and $z$ may acquire more linear pieces at nodes closer to the root. To see the runtime $T_{\mathrm{P}}$ of the parallel algorithm (Algorithm \ref{algorithm:threadi}) and the parallel speedup $S = T_{\mathrm{S}} / T_{\mathrm{P}}$ we start by estimating the number of nodes processed by thread~$p_0$ on the whole binomial tree. 

\begin{figure}[t]

\parbox[b]{.7\linewidth}{\includegraphics{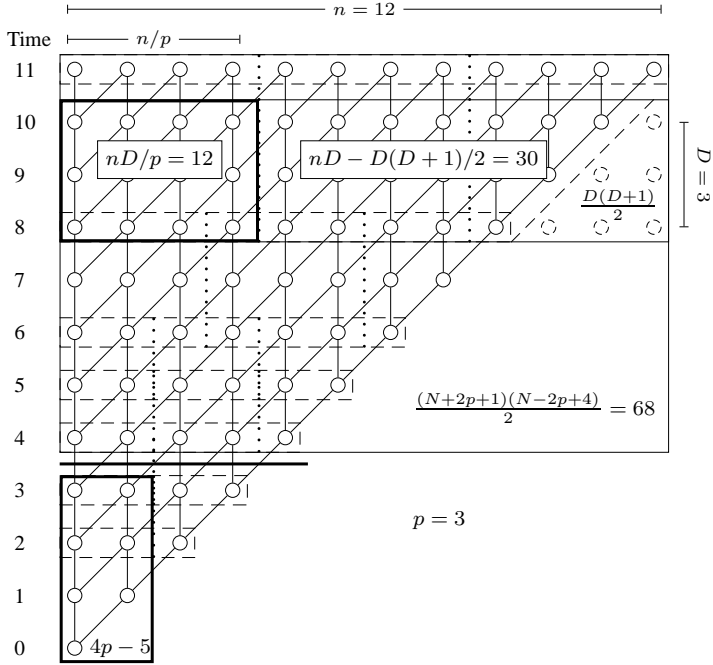}}\hfill
\parbox[b]{.3\linewidth}{

\caption{An estimation on the number of nodes processed by thread $p_0$.}
\label{figure:timeAnalysis}
}
\end{figure}

Generally, in a round whose base level is $B$ and has $n$ nodes, all these $p$ threads work in parallel on $D$ levels of the tree, from level $B - 1$ to $B - D$. According to the algorithm, the nodes within these $D$ levels will be divided into $p$ blocks, and the number of nodes assigned to thread $p_0$ is $nD/p$. The total number of nodes within these $D$ levels, assuming $n$ is an integral multiple of $p$, is $nD - \frac{D(D + 1)}{2}$. (See Fig. \ref{figure:timeAnalysis} for an example.) So the fraction done by thread $p_0$ is $nD/p$ divided by $nD - \frac{D(D + 1)}{2}$, which is $\frac{2}{p(2 - (D + 1)/n)}$. For large $n$ and relatively small $D$, we can assume that $(D + 1)/n \approx 0$, and, therefore, the fraction processed by thread $p_0$ is approximately $1/p$. This roughly applies to the part of the tree from the leaf level ($t = N + 1$) to the level where $t = 2p - 2$ ($p > 1$), because beyond this level further down the tree the number of processors needed will decrease. The total number of nodes in the tree from level $t = N + 1$ ($N + 2$ nodes) to level $t = 2p-2$ ($2p-1$ nodes) is $(N+2p+1)(N-2p+4)/2$, of which the number processed by thread $p_0$ is $(N+2p+1)(N-2p+4)/2p$. For the levels beyond $t=2p-2$, because thread $p_0$ will always have 2 nodes to process except at level $t=0$, the total number processed by $p_0$ will be $4p-5$. Therefore, for the whole binomial tree from $t=0$ to $t=N+1$, the total number of nodes processed by $p_0$ is $\frac{(N+2p+1)(N-2p+4)}{2p} + 4p-5$. If we assume $N \gg 2p$, then $\frac{(N+2p+1)(N-2p+4)}{2p} + 4p-5 \approx N^2/2p$.

To verify the validity of this estimation we have compared this estimated number with the actual counts obtained from several executions of the parallel algorithm. The data are summarised in Table \ref{table:verify}. The error rates are calculated and reported as well, from which it can be seen that the estimation is very close to the actual count in all the cases. For a fixed $p$ and $L$ ($D = \min(L, \lfloor (n+1) / p\rfloor - 1)$, jointly determined by $L$, $p$ and $n$), as the number $N$ increases the error rate decreases. This also is in-line with our analysis.

\begin{table}[h]
\caption{A comparison between $N^2/2p$ and the actual number of nodes processed by thread $p_0$ when $L = 5$. The fraction part of $N^2/2p$ is omitted.}
\label{table:verify}
\centering
\begin{footnotesize}
\begin{tabularx}{\linewidth}{Xlcrlcrlcr}
\hline

{\rule[-1mm]{0mm}{4mm}$p$} & \multicolumn{3}{c}{{\rule[-1mm]{0mm}{4mm}$N = 1200$}} & \multicolumn{3}{c}{{\rule[-1mm]{0mm}{4mm}$N = 1350$}} & \multicolumn{3}{c}{{\rule[-1mm]{0mm}{4mm}$N = 1500$}}\\

\cline{2-4} \cline{5-7}  \cline{8-10} \\ [-0.2cm]
&  {\rule[-0.5mm]{0mm}{1mm}Actual} & {\rule[-0.5mm]{0mm}{1mm}$N^2/2p$} & {\rule[-0.5mm]{0mm}{1mm}Error} & {\rule[-0.5mm]{0mm}{1mm}Actual} & {\rule[-0.5mm]{0mm}{1mm}$N^2/2p$} & {\rule[-0.5mm]{0mm}{1mm}Error} & {\rule[-0.5mm]{0mm}{1mm}Actual} & {\rule[-0.5mm]{0mm}{1mm}$N^2/2p$} & {\rule[-0.5mm]{0mm}{1mm}Error}\\  

\hline \\[-0.2cm]

2 & \sepnum{.}{,}{}{362999} &  \sepnum{.}{,}{}{360000} & -0.83\% & \sepnum{.}{,}{}{458999} &  \sepnum{.}{,}{}{455625} & -0.74\% & \sepnum{.}{,}{}{566249} &  \sepnum{.}{,}{}{562500} & -0.66\% \\[0.2cm]

4 & \sepnum{.}{,}{}{181198} &  \sepnum{.}{,}{}{180000} & -0.66\% & \sepnum{.}{,}{}{229161} &  \sepnum{.}{,}{}{227812} & -0.59\% & \sepnum{.}{,}{}{282748} &  \sepnum{.}{,}{}{281250} & -0.53\% \\[0.2cm]

8 & \sepnum{.}{,}{}{90311} &  \sepnum{.}{,}{}{90000} & -0.34\% & \sepnum{.}{,}{}{114255} &  \sepnum{.}{,}{}{113906} & -0.31\% & \sepnum{.}{,}{}{141008} &  \sepnum{.}{,}{}{140625} & -0.27\%\\

\hline
\end{tabularx}
\end{footnotesize}
\end{table}

Now since thread $p_0$ processes about $N^2/2p$ nodes, and the total number of nodes in a recombining binomial tree (from $t=0$ to $t=N+1$) is $(N+3)(N+2)/2 \approx N^2/2$, so the time required by $p_0$ is roughly $1/p$ of the sequential runtime. The sequential runtime $T_{\mathrm{S}}$ is $T_{\mathrm{S}} = \Oh(N^k)$ for some $k \ge 2$, and so the parallel runtime $T_{\mathrm{P}}$ is $T_{\mathrm{P}} = T_{\mathrm{S}} / p = \Oh(N^k) / p = \Oh(N^k / p)$.  The parallel speedup $S$ is therefore $S = T_{\mathrm{S}} / T_{\mathrm{P}} = \Oh(p)$, proportional to the number $p$ of processors used. So we can conclude by this analysis that the proposed parallel algorithm is cost-optimal in that $pT_{\mathrm{P}} = \Oh(N^k)$ having the same asymptotic growth rate as the sequential algorithms.

\section{Experimental results}
\label{section:testing}
\begin{figure}[t]

\parbox[b]{.57\linewidth}{\includegraphics{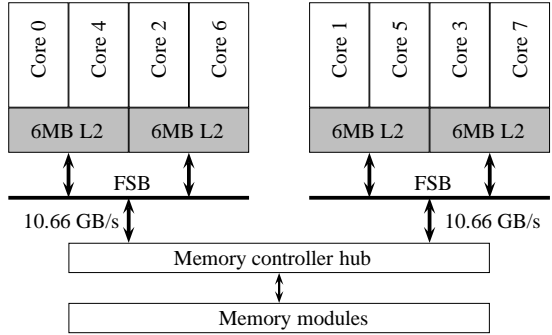}}\hfill
\parbox[b]{.43\linewidth}{

\caption{The parallel machine used in the tests.}
\label{figure:dualE5405}
}
\end{figure}

The parallel pricing algorithm was implemented in C/C++ via POSIX Threads, and was tested on a machine with dual sockets $\times$ quad-core Intel Xeon 2.0GHz E5405 running 8 processors in total (Fig. \ref{figure:dualE5405}). The source code was compiled by Intel C/C++ compiler icpc 12.0 for Linux. The testing machine was running Ubuntu Linux 10.10 64-bit version. The POSIX thread library used was NPTL (native POSIX thread library) 2.12.1.

To verify the correctness of the parallel algorithm we computed the ask and bid prices for the same American put option and the American bull spread described in Examples 5.1 and 5.2, respectively, in \cite{american-alet-2009}. In the American put example, the parameter values were $T=0.25$, $\sigma=0.2$, $R=0.1$, $S_0=100$, $K=100$, $N$ varied from 20 to 1000 and $k$ from 0 to 0.02. The American bull spread consists of a long call with $K = 95$ and a short call with $K = 105$, and is assumed to be settled in cash, with payoff process $(S_t - 95)^+ - (S_t - 105)^+$. In all the cases the parallel implementation produced exactly the same figures as reported in Table 1 and Table 2 in \cite{american-alet-2009}.

\begin{figure}

\parbox[b]{.74\linewidth}{\includegraphics{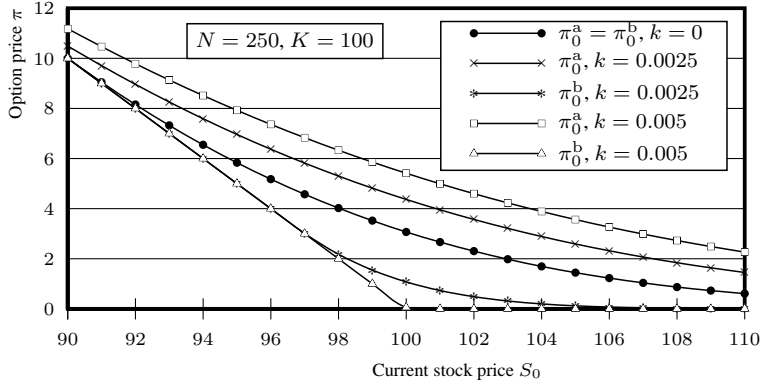}}\hfill
\parbox[b]{.26\linewidth}{\caption{Ask and bid price curves under different transaction cost rates.}
\label{figure:curves}
}

\end{figure}

To see the effect that proportional transaction costs have on option prices, we computed the prices for the same American put option (with $K = 100$) but with $S_0$ varying from 90 to 110 under three rates $k_0 = 0$, $k_1 = 0.25\%$ and $k_2 = 0.5\%$. The curves of the option prices $\pi_k^{\mathrm{a}}$ and $\pi_k^{\mathrm{b}}$ are plotted in Fig. \ref{figure:curves}, where it can be seen that for any fixed $S_0$ we have  $\pi_{k_2}^{\mathrm{b}} \le \pi_{k_1}^{\mathrm{b}} \le \pi_{k_0}^{\mathrm{a}} = \pi_{k_0}^{\mathrm{b}} < \pi_{k_1}^{\mathrm{a}} < \pi_{k_2}^{\mathrm{a}}$. Note that the larger the transaction cost rate $k$ the greater the ask-bid spread of the option price.

To test the performance of the parallel algorithm against an optimised implementation of the sequential algorithms we performed two additional sets of tests where $k$ was fixed to 0.005 for the American put option and to 0.01 for the bull spread, $N$ varied from 450 to 1500, and $p$ from 2 to 8. The runtimes and speedups are reported in Table \ref{table:performance}. All the times were wall-clock times measured in milliseconds (ms).

\begin{table}[!t]
\caption{Runtimes and speedups from the parallel performance tests.}
\label{table:performance}
\centering
\begin{footnotesize}
\begin{tabularx}{\linewidth}{Xlccccccr}
\hline

{\rule[-1mm]{0mm}{4mm}$p$ $|$ $S$} & {\rule[-1mm]{0mm}{4mm}$N\!=\!450$} & {\rule[-1mm]{0mm}{4mm}$N\!=\!600$} & {\rule[-1mm]{0mm}{4mm}$N\!=\!750$} & {\rule[-1mm]{0mm}{4mm}$N\!=\!900$} &  {\rule[-1mm]{0mm}{4mm}$N\!=\!1050$} & {\rule[-1mm]{0mm}{4mm}$N\!=\!1200$} & {\rule[-1mm]{0mm}{4mm}$N\!=\!1350$} & {\rule[-1mm]{0mm}{4mm}$N\!=\!1500$}\\

\hline \\[-0.1cm]

\multicolumn{9}{l}{American put $k=0.5\%$, $K=100$, $S_0 = 100$, $T = 0.25$, $R = 0.1$, $\sigma = 0.2$, $L = 5$} \\[0.3cm]

Serial &  181.0&	325.1&	498.6&	714.0&	979.3&	1302.1&	1608.4&	1983.7 \\[0.1cm]

$p=2$ & 128.7&	228.5&	348.7&	510.1&	679.9&	892.1&	1128.2&	1405.5 \\ [0.1cm]

$S$ & 1.41&	1.42&	1.43&	1.40&	1.44&	1.46&	1.43&	1.41 \\[0.1cm]

$p = 3$ & 90.4&	158.2&	241.8&	339.4&	468.8&	617.1&	765.0&	944.7\\[0.1cm]

$S$ & 2.00&	2.06&	2.06&	2.10&	2.09&	2.11&	2.10&	2.10\\[0.1cm]

$p = 4$ & 68.6&	121.6&	184.0&	268.7&	355.1&	469.7&	581.2&	724.8\\[0.1cm]

$S$ & 2.64&	2.67&	2.71&	2.66&	2.76&	2.77&	2.77&	2.74\\[0.1cm]

$p=5$ & 57.2&	96.8&	151.7&	213.0&	286.6&	374.7&	466.9&	583.7\\ [0.1cm]

$S$ & 3.17&	3.36&	3.29&	3.35&	3.42&	3.47&	3.44&	3.40\\[0.1cm]

$p = 6$ & 50.0&	83.7&	132.2&	187.2&	245.9&	313.9&	398.7&	493.2\\[0.1cm]

$S$ & 3.62&	3.88&	3.77&	3.81&	3.98&	4.15&	4.03&	4.02\\[0.1cm]

$p = 7$ & 43.5&	74.4&	115.3&	165.0&	214.7&	281.5&	363.8&	428.3\\[0.1cm]

$S$ & 4.16&	4.37&	4.33&	4.33&	4.56&	4.63&	4.42&	4.63\\[0.1cm]

$p = 8$ & 40.4&	67.3&	102.8&	142.7&	189.4&	248.6&	312.5&	376.8\\[0.1cm]

$S$ & 4.48&	4.83&	4.85&	5.00&	5.17&	5.24&	5.15&	5.26\\[0.3cm]

\multicolumn{9}{l}{American bull spread $k=1\%$, $\xi_t = (S_t\! -\! 95)^+\! -\! (S_t\! -\! 105)^+$, $T = 0.25$, $R = 0.1$, $\sigma = 0.2$, $L = 5$} \\[0.3cm]

Serial & 185.5&	327.9&	510.7&	731.2&	989.7&	1291.4&	1625.9&	2005.2\\[0.1cm]

$p=2$ &  133.1&	233.9&	365.2&	522.2&	699.5&	906.7&	1152.1&	1422.8\\ [0.1cm]

$S$ &  1.39&	1.40&	1.40&	1.40&	1.41&	1.42&	1.41&	1.41\\[0.1cm]

$p = 3$ &  95.9&	164.6&	254.7&	360.6&	486.2&	624.5&	781.8&	992.2\\[0.1cm]

$S$ & 1.93&	1.99&	2.01&	2.03&	2.04&	2.07&	2.08&	2.02\\[0.1cm]

$p = 4$ & 76.3&	130.9&	203.1&	279.7&	369.8&	474.6&	596.3&	734.3\\[0.1cm]

$S$ & 2.43&	2.50&	2.51&	2.61&	2.68&	2.72&	2.73&	2.73\\[0.1cm]

$p=5$ &  64.1&	106.6&	166.5&	229.3&	305.0&	393.0&	498.6&	622.5\\ [0.1cm]

$S$ &  2.89&	3.07&	3.07&	3.19&	3.24&	3.29&	3.26&	3.22\\[0.1cm]

$p = 6$ & 56.2&	91.4&	142.6&	197.7&	261.7&	334.6&	419.4&	510.0\\[0.1cm]

$S$ & 3.30&	3.59&	3.58&	3.70&	3.78&	3.86&	3.88&	3.93\\[0.1cm]

$p = 7$ & 48.2&	80.9&	124.9&	171.7&	228.9&	291.9&	364.5&	444.4\\[0.1cm]

$S$ & 3.85&	4.05&	4.09&	4.26&	4.32&	4.42&	4.46&	4.51\\[0.1cm]

$p = 8$ & 47.3&	79.6&	121.5&	167.6&	215.1&	273.6&	337.7&	403.6\\[0.1cm]

$S$ & 3.92&	4.12&	4.20&	4.36&	4.60&	4.72&	4.81&	4.97\\[0.1cm]

\hline
\end{tabularx}
\end{footnotesize}
\end{table}

\begin{figure}[t]
\centering

\parbox[t]{.74\linewidth}{\subfloat[Runtimes of the sequential program and the parallel one on 8 processors.]{\includegraphics{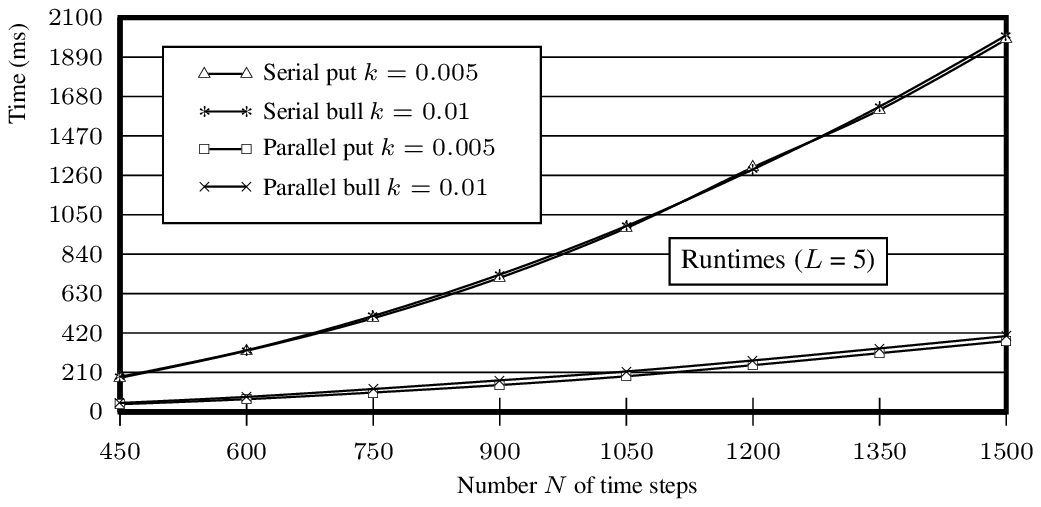}\label{subfigure:runtimes}} \\

\subfloat[Parallel speedups and efficiencies for $N=1500$.]{\includegraphics{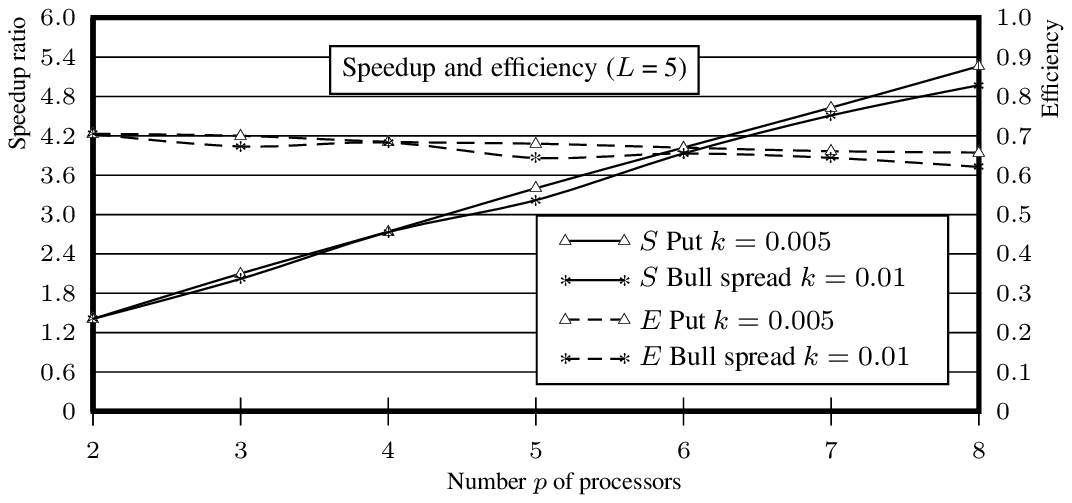}\label{subfigure:speedups1500Steps}}}\hfill
\parbox[t]{.26\linewidth}{

\caption{Plots derived from the experimental results.}
\label{figure:testing}
}
\end{figure}

Moreover, the serial and parallel runtimes when $p = 8$ and the parallel speedups when $N = 1500$ are plotted in Fig. \ref{figure:testing}\subref{subfigure:runtimes} and  Fig. \ref{figure:testing}\subref{subfigure:speedups1500Steps}, respectively. The speedup curves are very close to straight lines and this supports our analysis that the parallel speedup $S$ is proportional to $p$. Tests for other values of $L$ were performed in which very close results were found.

From the speedup ratios we calculated the parallel efficiency $E = S/p$. The analysis indicates that $S = \Oh(p)$, so $E = S / p = \Oh(p) / p = \Oh(1)$, which means that the efficiency of this parallel algorithm should stay the same no matter how many processors are used. However, in practice, because the synchronisation cost grows as the number $p$ increases, we can expect that the efficiency will decay as more processors are used. The efficiency data are plotted as dashed curves in Fig. \ref{figure:testing}\subref{subfigure:speedups1500Steps}, where it can be seen that the efficiency diminishes only slightly as $p$ increases.

\section{Conclusion}
\label{section:conclusion}
We have presented a parallel algorithm (based on the sequential pricing algorithms proposed in \cite{american-alet-2009}) that computes the ask and bid prices of American options under proportional transaction costs, and a multi-threaded implementation of the algorithm. Using $p$ processors, the algorithm partitions a recombining binomial tree into multi-level blocks. The whole computation, starting from the leaf nodes and working backwards to the root of the tree, is divided into rounds, where in each of these rounds, a block of nodes is further partitioned and processed by multiple processors. Before the start of the next round the workload of each processor (thread) is adjusted according to the number of nodes at the next base level. The applicability of the partition method and the associated synchronisation scheme is not restricted by the values of the parameters $N$ (number of levels of the tree), $L$ (maximum number of levels processed in a round) or $p$ (number of processors). The parallel algorithm has theoretical speedup $S = \Oh(p)$ and is cost-optimal because $pT_{\mathrm{P}} = \Oh(p) \times \Oh(N^k / p) = \Oh(N^k)$ for some $k \ge 2$, which has the same asymptotic growth rate as the serial runtime $T_{\mathrm{S}}$. The parallel efficiency $E$ of the algorithm is $E = S/p = \Oh(1)$.

The implementation was tested for its correctness and performance. The results demonstrated reasonable speedups, e.g., 5.26 when $p=8$ and $N=1500$, against an optimised sequential program even for problems of small sizes. The performance of the implementation was in-line with the asymptotic analysis. It showed that, because no inter-computer communication was involved, the overhead of the parallelisation in the multi-threaded implementation was much reduced compared to some previous approaches based on message-passing interfaces. The parallel efficiency in the tests is seen to decay slightly as $p$ increases. 

For options whose lifetime is short (within months) a relatively small number (usually several thousand) of time steps may be sufficient to model the price changes of the underlying asset. To handle such cases the multi-threaded implementation on main-stream multi-core processors will normally be fast enough. But for pricing long-life options (expiring in years) where large numbers of time steps are needed the parallel algorithm may have to be adapted to more powerful platforms, such as many-core general purpose graphics units. We are also aiming at developing high-performance parallel algorithms for pricing multi-dimensional options under proportional transaction costs. Since for such cases a direct implementation of the maximum, minimum and gradient restriction operations on multi-dimensional structures could be difficult, we may have to resort to Monte Carlo simulations, which are easily parallelised, and run them on large-scale parallel architectures.

\appendix
\section*{APPENDIX}
\setcounter{section}{1}
\label{section:appendix}
The parallel binomial algorithm we have developed is not specific to the problem of pricing American options under proportional transaction costs. It can be easily adapted to other problems, such as the case of pricing American options without considering transaction costs. In such cases, for an $N$-step simulation the algorithm does not add an extra time step $t = N + 1$ to the binomial tree. The other difference is that without transaction costs, all the payoffs and the expectations become scalars, and so the maximum operations are performed on numbers rather than on piecewise linear functions. The runtime $T_{\mathrm{S}}$ of a sequential binomial American option pricing algorithm with no transaction costs is $T_{\mathrm{S}} = \Oh(N^2)$. So the parallel runtime $T_{\mathrm{P}} = \Oh(N^2 / p)$. The parallel speedup $S = \Oh(p)$, and the parallel efficiency $E = \Oh(1)$.

Without considering dividends and transaction costs the price of an American call option is the same with a European call option under the same conditions \cite{options-john-2009}, so we consider only an American put option. We tested on the 8-processor machine (Fig. \ref{figure:dualE5405}) the performance of the parallel algorithm (modified in the two aforementioned aspects) using an American put option with strike $K=100$ and where the parameters were $S_0 = 100$, $T = 3$, $\sigma = 0.3$ and $R = 0.06$. In the test the number $N$ of time steps grew from 5000 to 40000, and the number $p$ of processors from 2 to 8. All the numeric variables in the program were represented by 8-byte double-precision floats. The runtimes and the speedups against an optimised sequential program are reported in Table \ref{table:performance}. All the times were wall-clock times measured in milliseconds (ms). The computed price for the American put option was 13.906.

\begin{table}[!t]
\caption{Runtimes and speedups from the parallel performance tests -- without transaction costs.}
\label{table:performance-without}
\centering
\begin{footnotesize}
\begin{tabularx}{\linewidth}{Xp{0.8cm}lXXXXXXr}
\hline

{\rule[-1mm]{0mm}{4mm}$p$ $|$ $S$} & {\rule[-1mm]{0mm}{4mm}$N=$} & {\rule[-1mm]{0mm}{4mm}$5000$} & {\rule[-1mm]{0mm}{4mm}$10000$} & {\rule[-1mm]{0mm}{4mm}$15000$} & {\rule[-1mm]{0mm}{4mm}$20000$} &  {\rule[-1mm]{0mm}{4mm}$25000$} & {\rule[-1mm]{0mm}{4mm}$30000$} & {\rule[-1mm]{0mm}{4mm}$135000$} & {\rule[-1mm]{0mm}{4mm}$40000$}\\

\hline \\[-0.1cm]

\multicolumn{9}{l}{American put $K=100$, $S_0 = 100$, $T = 3$, $R = 0.06$, $\sigma = 0.3$, $L = 50$} \\[0.3cm]

Serial &  &38.92&	158.88&	358.78&	638.26&	997.05&	1436.24&	1955.09&	2553.47 \\[0.1cm]

$p=2$ & & 21.3&	74.4&	160.7&	279.7&	433.4&	629.2&	927.2&	1411.9\\ [0.1cm]

$S$ &  &1.83&	2.13&	2.23&	2.28&	2.30&	2.28&	2.11&	1.81\\[0.1cm]

$p = 3$ & & 16.6&	54.6&	115.1&	197.8&	302.3&	429.2&	578.2&	756.6\\[0.1cm]

$S$ & & 2.34&	2.91&	3.12&	3.23&	3.30&	3.35&	3.38&	3.37\\[0.1cm]

$p = 4$ & & 16.0&	47.1&	95.1&	159.4&	239.7&	337.2&	451.3&	581.9\\[0.1cm]

$S$ & & 2.43&	3.38&	3.77&	4.00&	4.16&	4.26&	4.33&	4.39\\[0.1cm]

$p=5$ & & 15.1&	42.0&	82.3&	136.4&	203.1&	284.2&	378.8&	509.7\\ [0.1cm]

$S$ & &2.57&	3.79&	4.36&	4.68&	4.91&	5.05&	5.16&	5.01\\[0.1cm]

$p = 6$ & &15.1&	41.1&	77.1&	124.7&	182.9&	252.1&	333.2&	436.5\\[0.1cm]

$S$ & & 2.57&	3.87&	4.65&	5.12&	5.45&	5.70&	5.87&	5.85\\[0.1cm]

$p = 7$ & & 15.1&	41.1&	74.8&	117.5&	169.6&	231.3&	302.7&	386.3\\[0.1cm]

$S$ & & 2.57&	3.87&	4.80&	5.43&	5.88&	6.21&	6.46&	6.61\\[0.1cm]

$p = 8$ & &15.1&	41.1&	74.6&	114.3&	162.1&	217.8&	283.0&	356.3\\[0.1cm]

$S$ & & 2.57&	3.87&	4.81&	5.58&	6.15&	6.59&	6.91&	7.17\\[0.3cm]

\hline
\end{tabularx}
\end{footnotesize}
\end{table}

\begin{figure}[t]
\centering

\parbox[t]{.76\linewidth}{\subfloat[Runtimes of the sequential program and the parallel one on 8 processors.]{\includegraphics{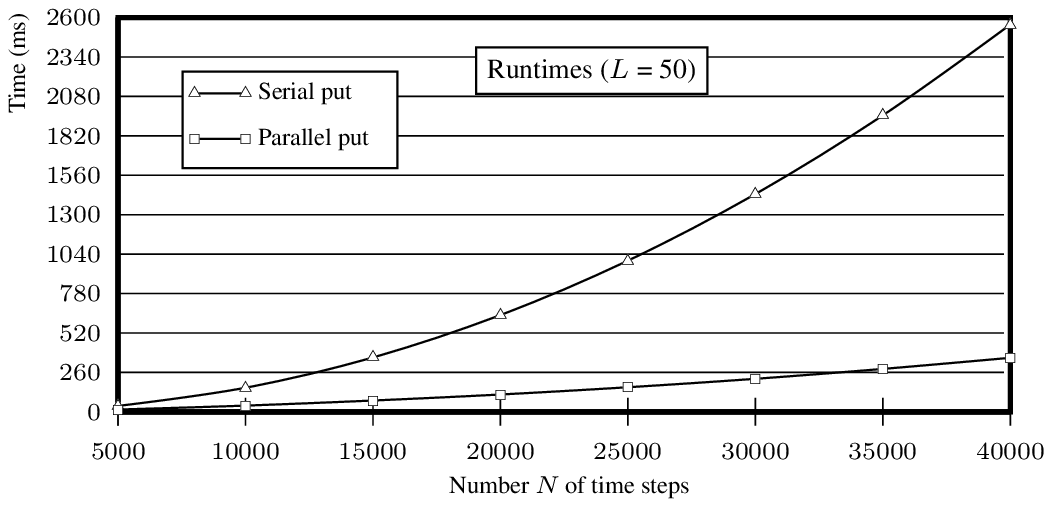}\label{subfigure:runtimes-without}} \\

\subfloat[Parallel speedups and efficiencies for $N=40000$.]{\includegraphics{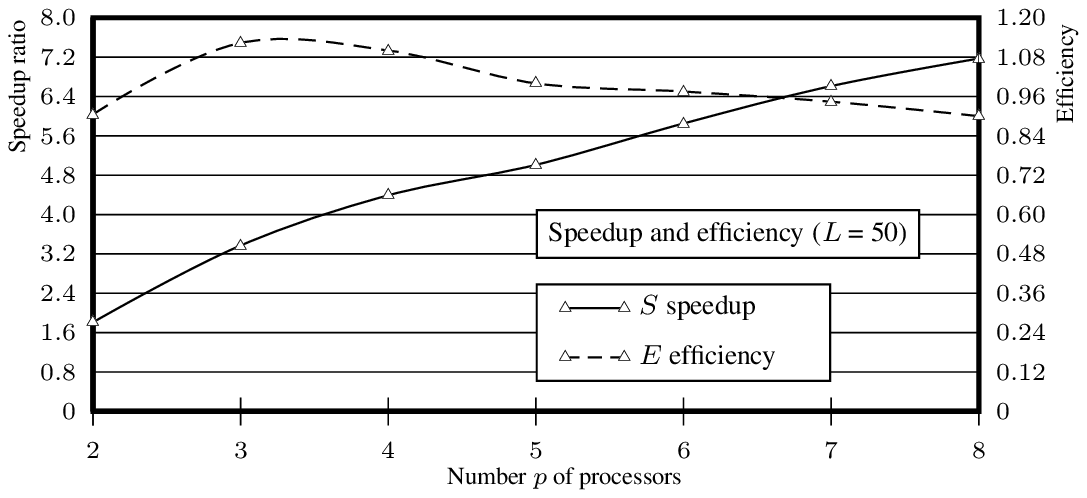}\label{subfigure:speedups40000Steps}}}\hfill
\parbox[b]{.24\linewidth}{

\caption{Plots derived from the performance tests on the American put option without transaction costs.}
\label{figure:testing-without}
}
\end{figure}

The serial and parallel runtimes when $p = 8$ and the parallel speedups when $N = 40000$ are plotted in Fig. \ref{figure:testing-without}\subref{subfigure:runtimes-without} and  Fig. \ref{figure:testing-without}\subref{subfigure:speedups40000Steps}, respectively. The parallel efficiencies were calculated from the speedups and plotted in Fig. \ref{figure:testing-without}\subref{subfigure:speedups40000Steps} as well.

From the results we observed super-linear speedups in several test cases, e.g., when $N = 30000$, $p = 3$ and the speedup $S = 3.35$. This was caused partly by the caching effect. The serial program can only use one of the four L2 caches (Fig. \ref{figure:dualE5405}), but the parallel program uses all the four. Moreover, the parallel program makes use of both the two FSBs, whereas the serial program uses only one. This also helps to increase the rate at which data is transferred between the main memory and the processors.

In all the tests parameter $L$ (the maximum number of levels being processed in a round) was set to 50, much increased from its value ($L = 5$) in the tests where transaction costs are present. The purpose of increasing its value was to reduce the number of times where the threads have to be synchronised, and therefore reduce the cost of the synchronisation. In the tests where transaction costs are considered, because the computation time was long enough relative to the synchronisation time, the performance was not as sensitive to the synchronisation overhead.  


\bibliographystyle{acmsmall}      
\bibliography{parallelbinomial-v4}   

\received{}{}{}



\end{document}